\documentclass[a4paper,twoside]{article}
\newcommand{\affil}[1]{$^{\rm #1}$}
\usepackage[authoryear]{natbib}
\bibpunct{(}{)}{;}{a}{}{,}
\usepackage{graphicx}
\date{} 
\title{\large\bf\flushleft Dynamical Influences of the Last Magellanic
Interaction On the Magellanic Clouds}
\author{\parbox{\textwidth}{\flushleft
\vspace{-0.5cm}
{\it Kenji Bekki\affil{A,C} and Masashi Chiba\affil{B}}\\
\vspace{0.4cm}
{\small \affil{A}\, 
School of Physics, University of New South Wales, Sydney 2052, Australia}\\
{\small \affil{B}\,Astronomical Institute,
Tohoku University, Sendai, 980-8578, Japan}\\
{\small \affil{C}\,Email: bekki@phys.unsw.edu.au}}}
\begin{document}
\twocolumn[
\begin{changemargin}{.8cm}{.5cm}
\begin{minipage}{.9\textwidth}
\vspace{-1cm}
\maketitle
%
%
\small{\bf Abstract:}
We investigate the present distributions of gas and young stars 
in the Large and Small Magellanic
Clouds (LMC and SMC) based on fully self-consistent numerical simulations
of the Clouds for the last $\sim 0.8$ Gyr.
Our principal results, which can be tested against observations,
are as follows.
The last dynamical and hydrodynamical
interaction  between the Clouds about $\sim 0.2$
Gyr ago can form the apparently
off-center bar and  peculiar  HI spirals of the LMC.
The present spatial distributions of young stars with ages
less than $\sim 20$  Myr in the LMC can be
significantly asymmetric and clumpy
owing to the interaction.
A small but non-negligible fraction of stellar and
gaseous components can be transferred from the SMC into the LMC
during the interaction to form diffuse halo components around the LMC.
The burst of star formation in the SMC  
can be synchronized with that of the LMC about 0.2 Gyr ago in some models.
New stars can form from gas in the SMC's tidal tails, one of which
can be observed as the Magellanic Bridge (MB).
The metallicity distribution function of new stars
in the MB has a peak of ${\rm [Fe/H]} \sim -0.8$, which is significantly
smaller than the stellar metallicity of the SMC. 
Based on these results, we discuss the origin
of 30 Doradus, the southern molecular ridge of the LMC,
the globular cluster ESO 121-SC03,  metal-poor inter-Cloud
stars within the MB, and giant HI holes of the LMC. \\



\medskip
\medskip
\end{minipage}
\end{changemargin}
]
\small

\section{Introduction}
The Large Magellanic Cloud (LMC) morphologically classified
as an irregular galaxy (Ir III-IV) is observed
to have a prominent off-center bar embedded within its flat disk component
(e.g., de Vaucouleurs \& Freeman 1972 hereafter dVF72; van den Bergh 2000, v00).
Recent observational studies based on
the Deep Near-Infrared Southern Sky Survey (DENIS)
and Two Micron All-Sky Survey (2MASS) have revealed
that the LMC has the off-center stellar bar in a significantly
elongated stellar disk (e.g., van der Marel 2001).
Influences of the off-center bar  on the evolution
of the LMC (e.g., the recent star formation  history)
have been discussed by several authors
(e.g., Gardiner et al. 1998; G98).

Such asymmetric structures seen in the LMC's  stellar components
have been revealed also in the gaseous ones
by multibeam HI survey of the LMC (e.g., Staveley-Smith et al. 2003, S03)
and by the first CO survey with NANTEN (Fukui et al. 1999). 
The  ``molecular ridge'' observed in the 
southern part of the LMC  has a peculiar distribution of molecular
clouds and is considered to be a possible formation cite of new
stars and clusters (Fukui et al. 1999; Kawamura et al. 2006).
Remarkably asymmetric and patchy distributions of star-forming complexes
and young populous clusters are also observed within the stellar
disk of the LMC (v00).
Although the observed asymmetric distributions
in variously different
populations (e.g.,  carbon stars, HI,  and star-forming regions)
are suggested to provide valuable  information on the LMC's evolution
interacting with the Small Magellanic Cloud (SMC) and the Galaxy
(e.g., Westerlund 1997),
it is unclear when and how these asymmetric
structures were formed in the dynamical history of the LMC.

The purpose of this paper  is thus to demonstrate,
for the first time, that the last dynamical interaction  between
the LMC and the SMC about 0.2 Gyr ago
can be responsible for the formation of
the off-center bar, peculiar spirals, and
asymmetrically distributed young stars in the LMC.
Although previous studies already pointed out
the importance of the Magellanic collision
in the formation of  the LMC's young star clusters
(e.g., Fujimoto \& Kumai  1997; Bekki et al. 2004a) and
a peculiar structure seen in the periphery of the LMC
(e.g., Kunkel et al. 1997),
the present study first explains
the origin  of asymmetric structures
in both stellar and  gaseous components of the LMC
in a self-consistent manner.

Although our main focus is
on the two-dimensional distributions of gas and stars
in the LMC, we also investigate 
the SMC's stellar and gaseous evolution that has not been 
investigated by previous numerical studies 
(e.g., Yoshizawa \& Noguchi 2003, YN03).
We particularly discuss
the origin of young, metal-poor stars observed in
the inter-Cloud regions such as the Magellanic Bridge (MB) 
(e.g., Rolleston et al. 1999).
We here focus on the essence of the physical mechanisms
(1) for the formation of asymmetric structure the LMC
and (2) for star formation in the MB
and accordingly will make more quantitative comparison of
the  simulation results
with latest observations in our forthcoming papers.

The plan of the paper is as follows: In the next section,
we describe our  numerical models  for
dynamical evolution of the LMC and the SMC 
interaction with each other and with the Galaxy.
In Section 3, we present the numerical results
on structural  properties and recent stat formation histories of  
the LMC and the SMC.
In Section 4, we discuss the derived numerical results in several
different contexts of the LMC's and the SMC's evolution, 
such as the origin of 30 Doradus in the LMC. 
We summarise our  conclusions in Section 5.

\begin{table*}
\begin{center}
\caption{Model parameters}\label{table1}
\begin{tabular}{ccccc}
\hline galaxy & $M_{\rm t}$ & $f_{\rm b}$ & $f_{\rm g}$ 
& $R_{\rm s}$     \\
\hline LMC & $2 \times 10^{10} {\rm M}_{\odot}$ & 
0.3 & 0.1 & 7.5 kpc  \\
\hline SMC & $3 \times 10^{9} {\rm M}_{\odot}$ & 
0.3 & 0.1 & 5.0 kpc   \\
\end{tabular}
\end{center}
\end{table*}

\section{The model}

Since fundamental methods and techniques of numerical simulations
on the evolution of the Clouds
are given  in our previous papers (Bekki \& Chiba 2005; BC05),
we briefly describe them here.
We investigate  dynamical and hydrodynamical evolution
and star formation histories of the Clouds from
0.8 Gyr ago ($T=-0.8$ Gyr)  to the present ($T=0$) by using
chemodynamical simulations
in which both the LMC and the SMC are modeled as
self-gravitating,  barred, and disky systems composed initially
of dark matter halos, stars (``old stars''),
and gas.
Total number of particles used for a galaxy is 120000
for all models in the present study.
We first determine the most
plausible and realistic orbits of the Clouds and then investigate the
evolution of the LMC using fully self-consistent GRAPE-SPH
simulations (Bekki \& Chiba 2006). In
determining the orbits, we adopt the same numerical method as those in
previous studies
(Murai \&  Fujimoto 1980; Gardiner \&  Noguchi 1996, GN96),
in which the equations of motion of the clouds
are integrated backward in time, from the present epoch until $\sim$ 0.8
Gyrs ago. The orbital evolution of the Clouds
in different models  is given in
BC05.

The two Clouds are modeled as self-gravitating disk galaxies
(with total masses of $M_{\rm t}$) 
embedded by massive dark matter halos (GN96, BC05).
The mass fraction of baryonic components (i.e., stellar and gaseous
ones) in a galaxy ($f_{\rm b}$) is set to be 0.3 both for
the LMC and for the SMC in the present study.  
The HI diameters of gas-rich galaxies are generally observed to be larger
than their optical disks (e.g., Broeils \& van Woerden 1994).
Guided by these observations, 
the gas disks ($R_{\rm g}$) of the Clouds are assumed to be at least
1.5 times larger than their stellar disks ($R_{\rm s}$) in the present models.
The initial $R_{\rm s}$ is  7.5 kpc for the LMC,
which means that both the stellar and gaseous disks 
are well within the tidal radius of the LMC if $R_{\rm g}/R_{\rm s} < 15$ kpc. 
The initial $R_{\rm s}$ is 5.0 kpc for the SMC which is similar
to that adopted in GN96.
The gas mass fraction ($f_{\rm g}$) is set to be 0.1 both
for the LMC and for the SMC.
The gas is converted into new field stars (``new stars'')
according to the Schmidt law with the observed threshold
gas density (Kennicutt 1998).
The adopted parameter values (e.g., $f_{\rm g}$) are summarized
in the Table 1.

The method to set up initial disks with stellar bars,
which are quite reasonable for the LMC and the SMC about $\sim 1$ Gyr
ago (BC05), is described as follows.
We first dynamically
relax the exponential disk composed purely of stars embedded
in the dark matter halo and thereby form a stellar bar (by invoking 
the global bar instability). Then we add an exponential  gas disk 
to the relaxed disk with the bar and give kinematical information
(e.g., circular velocities etc) to each of the gas particle.
Thus a disk with a stellar bar and an exponential gas disk
embedded in a massive dark matter halo can be set up for
the Clouds.

Although we obtain the results of models with different
masses (thus orbits) of the LMC ($M_{\rm L}$) and the SMC ($M_{\rm S}$)
for a reasonable range of their initial  masses,
we show the results of the {\it standard} model with
$M_{\rm L}=2.0 \times 10^{10} {\rm M}_{\odot}$ and
$M_{\rm s}=3.0 \times 10^{9} {\rm M}_{\odot}$.
This is firstly because the results of this model
can be compared with previous simulations with the same model
parameters (e.g., GN96)
and secondly because this model shows typical behaviors of
off-center bar formation.
$R_{\rm g}/R_{\rm s}$ is set to be 1.5 
for the LMC and 2.0 for the SMC in models.
The models with $R_{\rm g}/R_{\rm s}=1$ do not show
any significant differences in the spatial distributions
between gas and stars in the LMC. We do not intend
to show the result of these models, mainly because
they are not so consistent with observations.

We use the same coordinate system  $(X,Y,Z)$
(in units of kpc) as those used
in GN96 and BC05.
The adopted current positions
are $(-1.0,-40.8,-26.8)$  for the LMC
and $(13.6,-34.3,-39.8)$ for the SMC and
the adopted
current  Galactocentric radial velocity of the LMC (SMC)
is 80 (7) km s$^{-1}$.
Current velocities of the LMC
and the SMC in the
Galactic ($U$, $V$, $W$) coordinate
are assumed to be
(-5,-225,194) and  (40,-185,171) in units of km s$^{-1}$,
respectively.
The model for the disk configuration
(e.g., inclination angles) of the LMC (SMC)
is exactly the same as that in BC05 (GN96).
The initial inclinations of the Clouds with respect
to the above coordinate are set to be the same
as those adopted in GN96 (SMC) and BC05 (LMC).

Although the adopted orbital model is broadly consistent 
with previous ones that successfully explained
many observational results (GN96, BC05),
the orbital model of the SMC  is not so consistent with  
the recent observation  by Kallivayalil et al. (2006).
As shown in Fig 13 in Kallivayalil et al. (2006),
the LMC and the SMC
strongly interact with each other about 0.2 Gyr, if their
proper motion data  are considered in orbital calculations.
We thus suggest that as long as the last 1 Gyr evolution
is concerned, the models with orbital parameters suggested
by Kallivayalil et al. (2006) would show similar results shown
in the present paper.

\begin{figure}[h]
\begin{center}
\includegraphics[scale=0.4, angle=0]{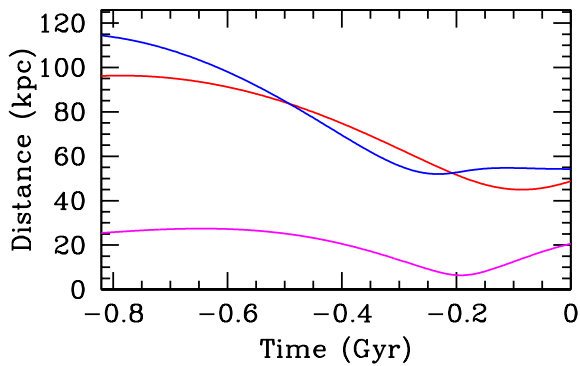}
\caption
{
Time evolution of distance between
the LMC and the SMC (magenta), the LMC and the Galaxy (red),
and the SMC and the Galaxy (blue) for the last 0.8 Gyr
in the standard model.
Note that the LMC-SMC distance becomes minimum (8kpc) about
0.2 Gyr ago.
}
\label{figexample}
\end{center}
\end{figure}

We derive the two-dimensional (2D) distributions of
the $B-$band surface brightness (${\mu}_{\rm B}$)
in the LMC's  stellar disk
from  the simulated mass
distributions of its old and new stars
by assuming that the stellar mass-to-light-ratio
($M/L_{\rm B}$)
is  4.0 for old stars and 0.16 for new ones
(Vazdekis et al. 1996).
In order to show more clearly the simulation
results in some figures,
we make the following coordinate transformation
for the projected mass distributions to see
more clearly the face-on morphology of the LMC:
Firstly the LMC disk is rotated by some degrees in
the $X$-$Z$ plane (i.e., about the $Y$-axis)
such that the major axis of the bar
is coincident with the $X$-axis,
then the disk is rotated by 180 degrees about
the $Z$-axis.
We investigate (1) the model without SMC (``SMC-less'' model)
and (2) the model in which the LMC interacts neither with
the Galaxy nor with the SMC (``isolated'' model)
so that we can more clearly understand the roles of SMC
in the LMC evolution.
We  also investigate the evolution of the Clouds from
$T=0$ Gyr to $T=+6$ Gyr (i.e., the next 6 Gyr)
to discuss the final dynamical fate of the Clouds.

Although we investigated many models with different orbits
depending  on the adopted initial velocities 
that are consistent with observations on proper motion
of the Clouds,
we here show the results of the standard model,
because this model enables us to describe clearly the 
essential roles of the last Magellanic interaction
in structural properties and recent star formation
histories of the Clouds in a convincing way.
We plan to discuss how numerical results depend on
a number of parameters of the Magellanic system
(e.g., orbits, masses  of the LMC and the SMC, and
parameters of star formation) in our forthcoming papers.
Since we mainly focus on two-dimensional (2D) distributions
of gas and stars  {\it in the present
Clouds},  
we do not intend to describe dynamical evolution
of the Clouds in the present study.
The details of dynamical evolution (e.g., morphological
evolution of the LMC) of the Magellanic system
have been already given in BC05.

\begin{figure*}
\begin{center}
\includegraphics[scale=0.9, angle=0]{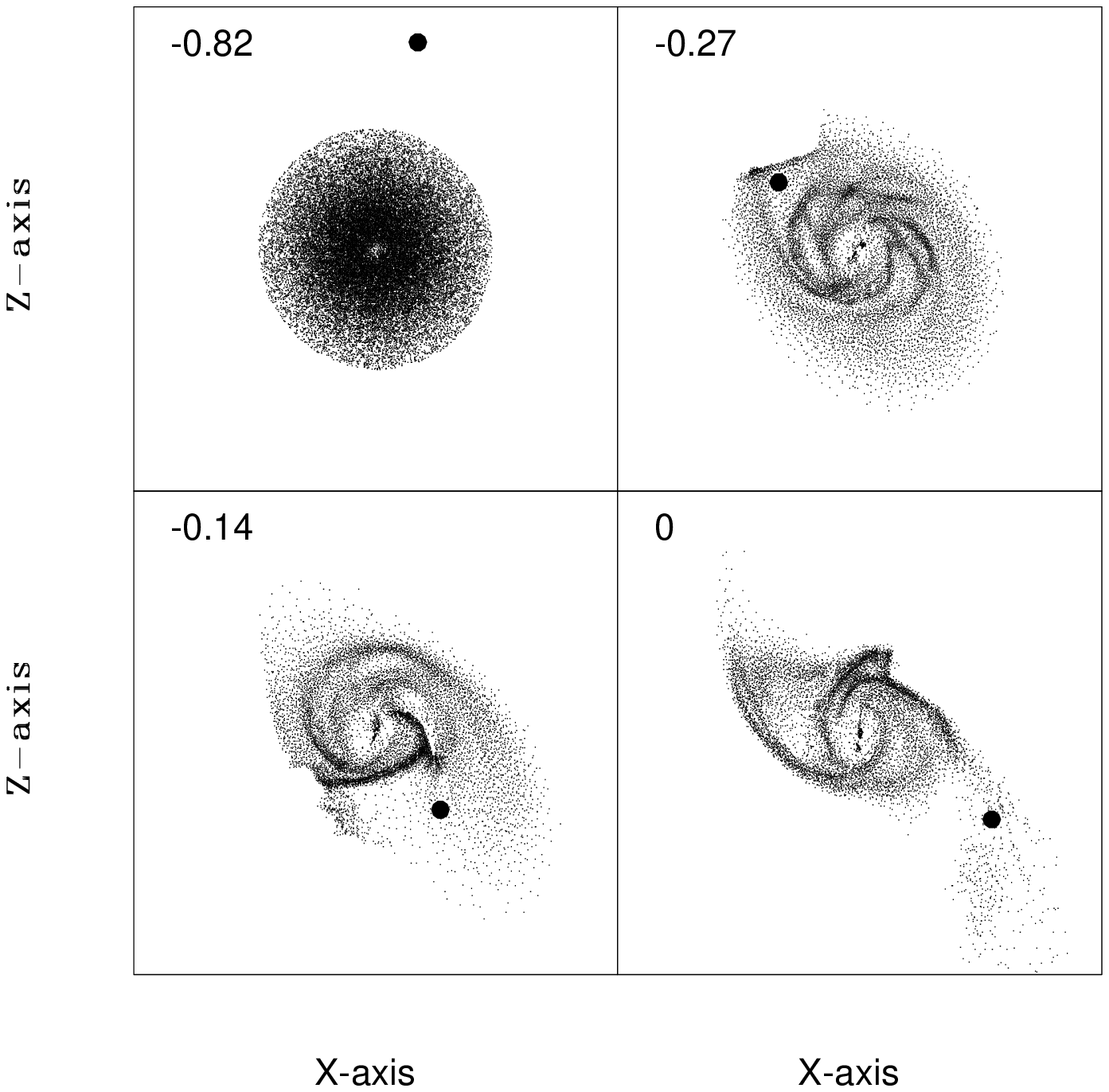}
\caption{
Distributions of the LMC's gas particles projected onto the
$X$-$Z$ plane at different epochs. 
For convenience,
the center of each frame is 
set to be coincident with the center of the LMC.
Each frame measures 45 kpc
and the number shows in the upper left corner of each frame represents
the time $T$ in units of Gyr. 
The gaseous evolution from 0.82 Gyr ago ($T=-0.82$) to the present
($T=0$) is shown in these four frames. 
For clarity, 
gas particles of the SMC are not shown
and the center of the SMC is shown by a big dot.
Total number of gas particles dramatically decreases owing
to gas consumption by star formation as time passes by.
Most gas in the central 3 kpc of the LMC's disk
is converted into new stars to form a stellar bar.
}
\label{figexample}
\end{center}
\end{figure*}

Figure 1 shows 
the orbital evolution of the LMC and the SMC with
respect to the Galaxy for the last $\sim$ 0.8 Gyrs
in the standard model.
Although the present model includes dynamical friction
between the LMC and the SMC (owing  to the fully
self-consistent model of the Magellanic system),
the orbital evolution is quite similar to that
in previous models without dynamical friction between
the Clouds (e.g., GN95; BC05).
Because of the small pericenter distance (=8 kpc) in the LMC-SMC
orbit,  the outer parts of the gas disks of the Clouds
collide with each other to form peculiar gaseous structures
in the Clouds for this model.
The pericenter distance of the LMC-SMC orbit,
which is a key parameter for the final 2D-distributions
of gas and stars in the Clouds,  strongly
depends on the initial velocities of the Clouds in the present models.
Thus it should be stressed here that it depends on
the adopted initial velocities of Clouds whether (and when) 
the Clouds collide with each other in the present models.

The pericenter distance of 8 kpc in the LMC-SMC orbit shown in Figure 1
about 0.2 Gyr ago
means that the tidal force from the LMC is  about 20 times stronger
than that from the Galaxy for the SMC.
This thus means  that the SMC
can be much more strongly influenced by the LMC-SMC interaction than
the SMC-Galaxy one. 
This LMC-SMC interaction can also significantly
influence stellar and gaseous evolution of the LMC and thus its recent
star formation history.
We thus discuss how the tidal interaction
controls the final (i.e., $T=0$) distributions of 
gas and young stars in the Clouds.


\begin{figure*}
\begin{center}
\includegraphics[scale=0.9, angle=0]{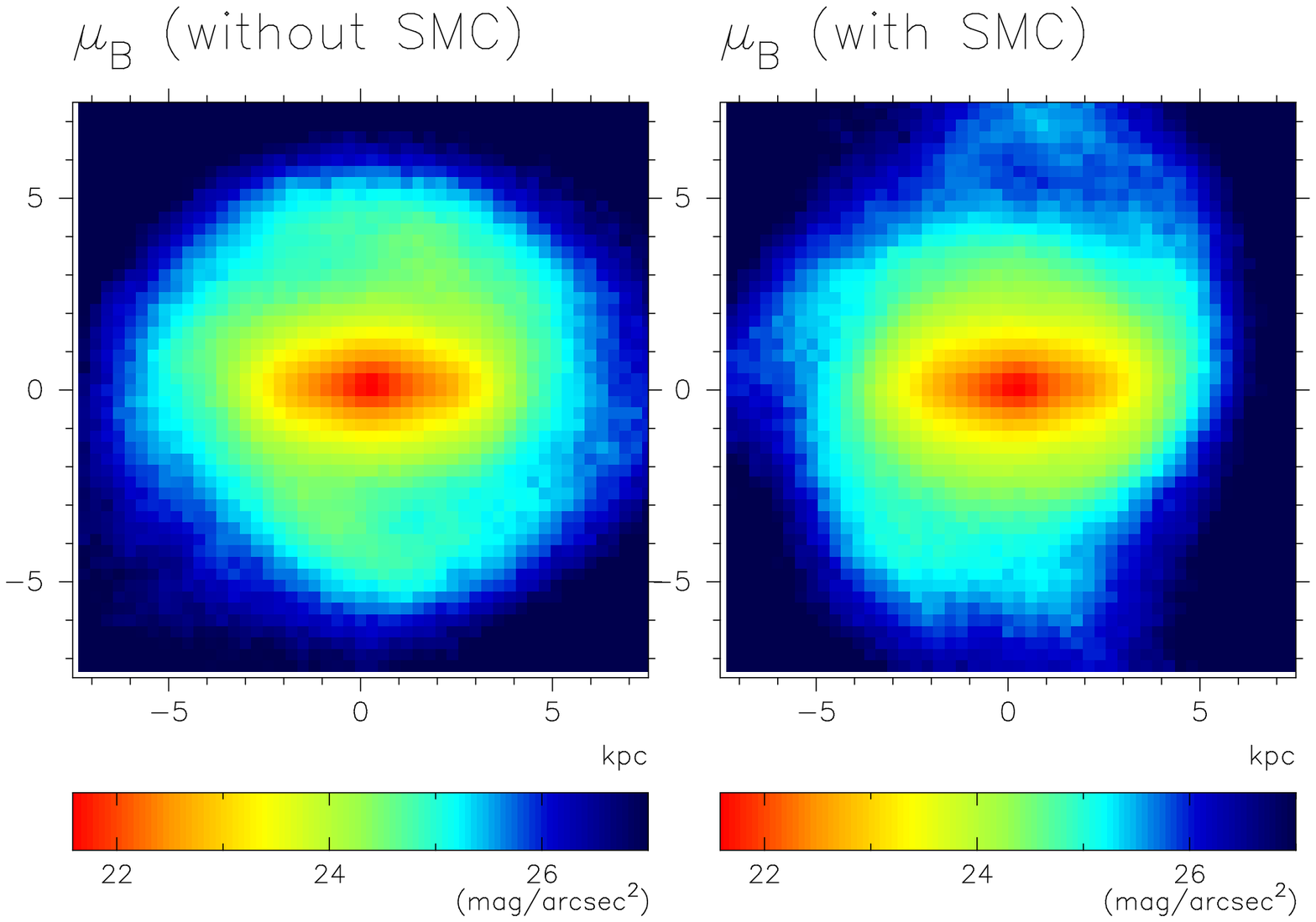}
\caption{
 The 2D distributions of 
the B-band surface brightness (${\mu}_{\rm B}$) 
of the LMC  at $T=0$ Gyr  projected onto the $X$-$Z$
plane (similar to the face-on view of the LMC)
for the SMC-less model (left) and the standard model with
the SMC (right). 
Here coordinate transformation is made
so that the results can be more readily compared
with the observations.
The center of the bar is set to be coincident with
the center of the frame for each model.
The 2D map consists of 2500 cells with the cell size of 0.3 kpc
and ${\mu}_{\rm B}$ is estimated for each cell.
Gaussian smoothing  is used in order to
derive smoothed ${\mu}_{\rm B}$ fields.
Note that although the SMC-less model shows a elliptic disk, 
the center of the stellar bar in the disk does not appear to be
coincident with the center of the outermost isophoto of the disk.
Note also that the disk in the standard model appears to have
a off-center bar.   
}
\label{figexample}
\end{center}
\end{figure*}

In the present study,
we show the results of the most important model
that should be compared with the corresponding observations,
though the results can depend strongly LMC's and SMC's orbits,
for which there are observational uncertainties. For example,
if we adopt the model with a much larger  pericenter distance ($>20$ kpc)
between the LMC and the SMC around 0.2 Gyr ago,  the asymmetric
appearance of the LMC becomes less remarkable and
the total amount of gas and stars stripped from the SMC
becomes also small.  We do not intend to discuss the
models with variously different parameters in the present
study, main because the results of most of these models
are not so consistent with observations.
It should be noted here that ``the isolated LMC model''
without interaction (of the LMC)
with the Galaxy and the SMC does not show any remarkable
asymmetric spiral arms seen in the model described in the
present study (e.g., Figure 2). This confirms that
the simulated asymmetric spiral arms are not due to numerical
artifacts of initial gaseous disks adopted in the present study.

\section{Results}
\subsection{The LMC}

Figure 2 shows the time evolution of the gaseous distribution
of the LMC's disk interacting with the SMC and the Galaxy.
Owing to the combined tidal fields of the SMC and the Galaxy,
the LMC's outer gas disk is strongly disturbed to form 
peculiar gaseous arms in the disk ($T=-0.27$ Gyr).  
The gas disk also hydrodynamically interacts with
the SMC's gas disk during the pericenter passage of
the SMC ($T=-0.14$ Gyr) so that asymmetric gaseous
arms with high-densities can be developed in the disk.
A  
small fraction of the LMC's gas can be stripped and dispersed
into the region between the LMC and the SMC (i.e., inter-Cloud region)
during tidal interaction to form diffuse gaseous components 
in the inter-Cloud region. 
An elongated gas disk with a number of asymmetric, peculiar  spiral
arms is finally formed without the  stellar disk
being severely disturbed ($T=0$ Gyr).

 Figure 3 shows the 2D distributions of ${\mu}_{\rm B}$
of the LMC  
projected onto the $X$-$Z$ plane at $T=0$ Gyr
for the standard model and the SMC-less one.
Although the tidal  interaction between the LMC
and the Galaxy can form an elliptical
stellar disk,  no off-center bar can be formed in
the SMC-less model.
The standard model, on the other hand, shows 
a clear asymmetric distribution in the outer disk
with respect to the 
central bar, and accordingly the central bar appears
to be largely shifted from the  center of the disk.
The results of these two models clearly demonstrate
that the Magellanic interaction can be responsible
for the formation of the apparently off-center bar
embedded within the LMC's elliptic stellar disk.
The result of the SMC-less model is quantitatively
consistent with previous models without SMC evolution
(e.g., BC05).

It should be stressed that the simulated asymmetric 2D distributions
are due to dynamical interaction between the LMC and the SMC 
(and the Galaxy): they are not due to the collision between
the two gas disks.  The contribution of old stars is $~ 90$ \%  in mass and
$\sim 26$ \% in light owing to the adopted  $M/L = 4.0$ for old stars.
This means that the derived  asymmetric 2D distributions
are due largely to young stars formed in the asymmetric gas.

 As Figure 4 reveals,  showing the 2D distribution of the surface gas
density (${\mu}_{\rm g}$) of the standard model,
the Magellanic interaction results in the formation of a significantly
asymmetric gas distribution with higher gas densities in the left
side of the stellar bar.
Two gaseous arms appear to emerge from above the left end of the bar
with the one extending to the direction of $(X,Z)\approx (-7,0)$ kpc
and the other extending to the direction  of $(X,Z) \approx (-2,-7)$ kpc.
The latter gaseous  arm is furthermore connected to the Magellanic
gaseous bridge in the inter-Magellanic region.
Owing to the mass-transfer from
the SMC into the LMC during the Magellanic interaction,
about 8\% of the LMC's gaseous componets  at $T=0$ Gyr
were initially within the SMC.
These transferred gas might  well form either the diffuse halo gas
around the LMC (S03) or kinematically odd components
such as the ``L-component''
(Luks \&  Rohlfs 1992).
Only $\sim 2$\% of old stellar components of the SMC
can be transferred into the central 7.5 kpc of the LMC.
Old stars and new stars formed from gas transferred from
the SMC can be discussed in terms of structurally and kinematically
odd stellar components of the LMC (e.g., Subramaniam \& Prabhu 2005).

It should be stressed here that the simulated gas distributions 
of the LMC do not show  remarkable lack of gas within the 
stellar bar: observations clearly show HI deficiency within the
off-center stellar bar (e.g., Staveley-Smith et al. 2003).
A possible  reasons for
the apparent failure of the models
is that  the present simulations  can not
model so realistically the conversion from HI to molecular clouds 
and to field stars and star clusters in the central region
of the LMC: star clusters and molecular gas are observed
in the bar region of the LMC.

\begin{figure}[h]
\begin{center}
\includegraphics[scale=0.8, angle=0]{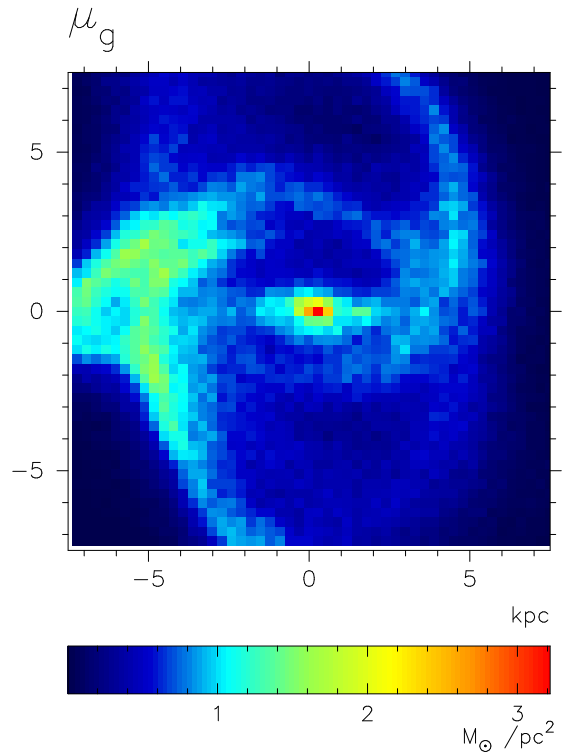}
\caption{
The 2D distribution of
the surface gas density  (${\mu}_{\rm g}$)
of the LMC  at $T=0$ Gyr projected onto the $X$-$Z$ plane
for the standard model.
This figure is a zoomed version of the last panel in Figure 2.
}
\label{figexample}
\end{center}
\end{figure}

\begin{figure}[h]
\begin{center}
\includegraphics[scale=0.8, angle=0]{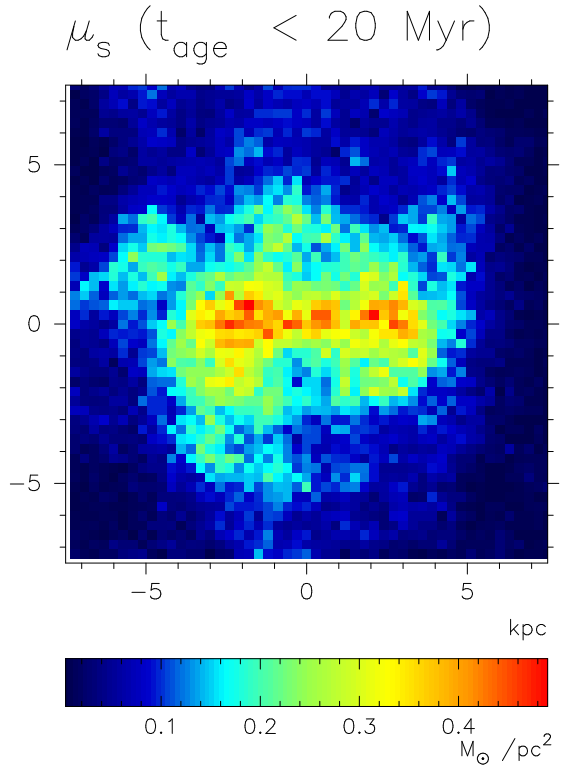}
\caption{
The 2D distribution of
the stellar surface  density  (${\mu}_{\rm s}$)
of the LMC  at $T=0$ Gyr projected onto the $X$-$Z$ plane
for the standard model.
Here the distribution is derived for very young stars
with the ages ($t_{\rm age}$) less than 20 Myr.
Note that 
there is an interesting peak just above the left
edge (or eastern edge) of the bar.
}
\label{figexample}
\end{center}
\end{figure}

 Figure 5 shows that the 2D distribution (${\mu}_{\rm s}$)
of young stars with
the ages ($t_{\rm age}$) less than 20 Myr is quite irregular
and patchy,  in particular, in the left half of the disk.
The high-density regions of young stars can be seen within
the stellar bar, where gas can be compressed into new stars
owing to dynamical action of the bar.
These irregular distributions of new stars can not be clearly
seen in the ${\mu}_{\rm s}$ distribution for all new stars with
$t_{\rm age}<200$ Myr, which suggests that
the LMC collision is responsible only for the asymmetric distribution
of very young stars ($t_{\rm age}<20$ Myr).
The location of a clumpy distribution
just above the left end of
the bar (i.e., $(X,Z) \approx (-5,2)$ kpc)
is coincident with
the high-density region of the gaseous arm connected to the
MB. This result is discussed in terms of 30 Doradus formation
later in this paper.

\begin{figure}[h]
\begin{center}
\includegraphics[scale=0.8, angle=0]{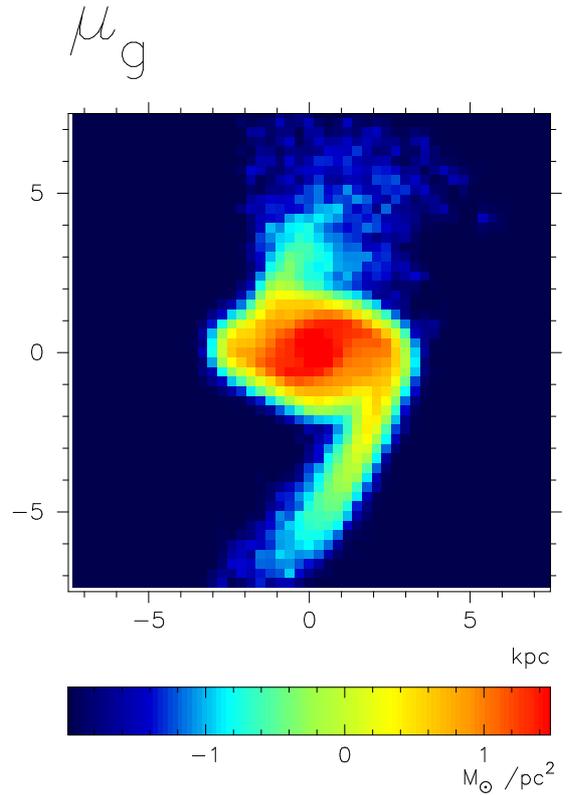}
\caption{
The projected distribution of smoothed (column) gas densities of the SMC about
0.14 Gyr ago (i.e., 60 Myr after the last Magellanic interaction).
The smoothed densities (${\mu}_{\rm g}$) are give in a logarithmic scale. 
The lower tidal tail with a higher gas density is the forming MB.
}
\label{figexample}
\end{center}
\end{figure}

\begin{figure}[h]
\begin{center}
\includegraphics[scale=0.4, angle=0]{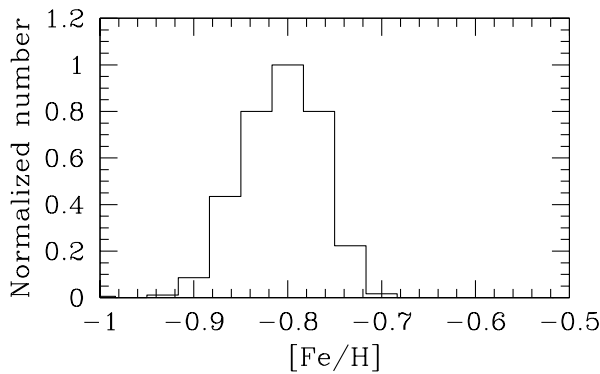}
\caption{
The metallicity distribution function of new stars formed in
the MB (normalized to the maximum number of stars in the  
metallicity bins). Note that the peak metallicity is around
${\rm [Fe/H]} = -0.8$, which is significantly smaller
than the central stellar metallicity of the simulated SMC.
}
\label{figexample}
\end{center}
\end{figure}

 The star formation rate averaged over the entire region
of the LMC in the standard model is sharply increased 
by a factor of 6 at $T=-0.2$ Gyr 
in comparison with the isolated model
with the star formation rate being almost constant  
$0.5 {\rm M}_{\odot}$ yr$^{-1}$ during the last 0.8 Gyr evolution.
Since previous (and the present study)
showed that the SMC shows a burst of star formation about 0.2 Gyr
ago in their simulations (YN03), 
the sharp burst of star formation derived above for the LMC implies
that (moderately strong)  bursts of star formation  can be synchronized
in the LMC and the SMC about 0.2 Gyr ago. 
This synchronized burst of star formation 
in the Clouds   
through the LMC-SMC tidal interaction 
can be proved by
the observed age distributions of young star clusters in the Clouds
(e.g., Girardi et al 1995).

It should be noted here that Harris \& Zaritsky (2004)
and Zaritsky \& Harris (2004) 
found a peak in the
SMC's star formation history that coincided with close passage to the
Milky Way.
They however were unable to conclude whether the LMC-SMC interaction
has had any noticeable effect on the SMC's star formation history.
Therefore, the simulated burst of star formation about 0.2 Gyr 
in the SMC has not been observationally confirmed yet.

 We confirm that the Clouds can merge with each other to
form a single Magellanic cloud within several  Gyr
from the
present ($T=0$ Gyr) in {\it some } models, because
the SMC continues to sink into the LMC owing to dynamical
friction between the LMC's halo and the SMC.
The LMC is transformed into a disk galaxy with a S0-like morphology,
a bar,  a thick stellar disk, and rotational kinematics during merging.
The  newly formed ``Magellanic cloud''
also has a stellar halo formed mainly
from the tidal debris of the SMC and accordingly
the expected mean metallicity of the halo is as high
as ${\rm [Fe/H]} \sim -0.7$,  which is significantly
higher than that of the Galactic old stellar halo
(${\rm [Fe/H]} \sim -1.6$).
These results will be discussed by our forthcoming
papers in terms
of the origin of stellar halos in S0s.

\subsection{The Magellanic Bridge}

The outer part of the SMC's gas disk is 
strongly stretched during the LMC-SMC interaction
to form two tidal tails,
one of which is observed as the MB.
Since the HI gas disk is initially larger
than the stellar one in the SMC model
with $R_{\rm g}/R_{\rm s}=2$,
the MB is composed of metal-poor gas and stars
formed from the gas (i.e, almost no old stars originating
from the inner stellar disk in the MB).
About 25\% of the initial gas of the SMC is stripped to form
the MB, which means that the total mass in the MB
is about  $10^{8} {\rm M}_{\odot}$ 
for the SMC model with the gas mass fraction of 0.5.
The final gas mass in the MB depends on $R_{\rm g}/R_{\rm s}$
of the SMC in the sense that
a larger amount of gas can be distributed in the MB
for the models with larger $R_{\rm g}/R_{\rm s}$.

The LMC-SMC tidal interaction 
can also significantly change the recent star formation
histories
not only in the central region of the SMC's gas disk  but also in
its outer part (YN03).
Figure 6 shows that
the SMC's outer gas disk is strongly compressed by the interaction
so that gas densities along
the forming MB can exceed the threshold
gas density of star formation (i.e., $3 {\rm M}_{\odot}$ pc$^{-2}$).
New stars  therefore can form in the MB and some of the stars
are finally stripped to become inter-Cloud stars with
young ages with less than 200 Myrs. 
The spatial distribution 
of these simulated inter-Cloud stars
can be compared with that of blue stars observed
in the inter-Cloud region (e.g., Irwin et al. 1990; Rolleston et al. 1999).

Since the MB is formed from the outer gas, where the metallicity
is significantly smaller owing to the negative metallicity gradient
of the SMC,  
the metallicities of young stars in the MB
can be significantly small: the metal-poor stars are due to
the adopted metallicity gradient of the SMC's gas disk
(not due to the dynamical mixing during the interaction).
Figure 7 shows that
the metallicity distribution of new stars in the MB
has a peak of [Fe/H] $\sim$ $-0.8$,
which is about  0.2 dex smaller than
the central stellar metallicity of the simulated SMC.
Although this result is consistent with
the presence of young, metal-poor stars
in the MB,
the observed stars with  metallicities
0.5 dex lower than those of  SMC's stars (i.e., corresponding
to ${\rm [Fe/H]} \sim -1.1$)
can not be well reproduced.
Only 0.1\% of the gas mass of the SMC is converted into
new stars in the MB, which means that the mass ratio of
new stars to gas in the MB is $\sim 0.004$.
The present model thus provides a physical explanation for the
origin of the observed formation sites of new stars along
the MB
(Mizuno  et al. 2006).

Recently Harris (2006) has reported a possible evidence
for the MB without old field stars and suggested
that the MB was formed from purely gaseous components.
The present models, however, are not consistent with
this result in the sense that a minor fraction of
old disk stars can be stripped tidally
from the SMC and still  within the MB at the present time.
We will discuss this problem in a more extensive and quantitative
way in our future papers.

\section{Discussions}

\subsection{Formation of 30 Doradus}

30 Doradus and young clusters within and close to it
have been extensively discussed both theoretically
and observationally in the contexts of
triggered star formation by stellar feedback effects
and initial mass functions (IMFs) of star clusters
(e.g., Westerlund 1997; Selman et al. 1999;  
v00).
However it remains unclear  how giant molecular
cloud(s) can be converted into massive, new clusters
embedded in the HII region of 30 Doradus in the evolution
of the LMC.  
In particular, it is unclear how {\it galaxy-scale dynamics}
is important for the formation of massive gas clouds
that are progenitors of 30 Doradus. 
We here consider that
the location of 30 Doradus with respect to
HI gas and molecular gas provides a vital clue
to the formation process of 30 Doradus.

 30 Doradus with the central super-cluster R136 is observed
to be located just above the eastern end of the LMC's bar
(e.g., v00) and the peculiar HI arms
appear to emerge from above the eastern end of the bar
(See Figure 1 in S03).
Large scale mapping of molecular gas in the vicinity of 30 Doradus
in the LMC has recently discovered a strong concentration
of GMCs close to 30 Doradus
(Ott et al. 2006). 
Interestingly, the present simulations also show that
both gaseous arms with high densities and relatively
densely populated
regions of very young stars ($t_{\rm age}<$ 20 Myr)
can be seen above the left side of the LMC's  bar
(corresponding to the eastern edge of the bar).
This amazing coincidence between the observations and
the simulations implies that the origin of 30 Doradus
(and thus R136) could result from dynamical and hydrodynamical
interaction between the Clouds in the  last Magellanic interaction.

Thus the preset simulations suggest that the observed
unique location of 30 Doradus tells us
about 
the formation process  of 30 Doradus:
30 Doradus originates from GMCs that were developed in  high-density regions
of asymmetric gaseous spiral arms formed in the
last Magellanic interaction. 
Since the dynamical influences of galaxy interaction
can last for one rotation period (corresonding to
0.25 Gyr for the LMC),
the formation
of star clusters with ages less than 0.25 Gyr in the LMC
might well be triggered by the Magellanic interaction.
We accordingly suggest that 30 Doradus is
just one example of many star-forming regions triggered 
by  the interaction with the SMC (and the Galaxy) 
and the resultant large-scale
dynamical effects inside the LMC.

The distribution of star-forming region like  30 Doradus
is observed to be asymmetric,  irregular,  and patchy in the LMC.
The origin of such a  distribution of star-forming
regions and young massive stars has long been discussed 
in terms of different physical mechanisms, such as
stochastic self-propagating star formation (SSPSF; Feitzinger et al. 1981),
gas compression in a bow shock due to the Galactic warm gas
(de Boer et al. 1998), and dynamical perturbation of the
LMC's off-center bar (G98).
The simulated asymmetric distribution of young stars
in the LMC disk suggests that the Magellanic collision
can also play a role  in forming  
asymmetric distributions of star-forming regions and young stars.
It is thus our future study to understand the relative importance
of these mechanisms (e.g., SSPSF and the Magellanic collision)
in the formation of the asymmetric
distributions in the LMC.

\subsection{Off-center bars and dynamical influences of
dark matter subhalos}

Although barred Magellanic-type dwarfs
that appear to be pairs of galaxies
are not rare (e.g., Freeman 1984),
the existence of off-center bars is seen in systems that
are not obviously interacting (dVF72; G98).
Recent cosmological  numerical simulations
(e.g. Susa \& Umemura 2004) have demonstrated
that star formation
in low-mass systems with the masses of $\sim 10^8 {\rm M}_{\odot}$
can be  severely suppressed by the cosmic reionization
so that these systems can become galaxies that are not optically
observable (``dark galaxies'').
We accordingly suggest that
the formation of off-center bars in
Magellanic-type dwarfs with  apparently no optical companions
could be possibly due to
the unequal-mass galaxy collision
between them and their invisible companions:The presence
of off-center bars can be useful for proving missing satellites
through interaction with them.

\subsection{Origin of  ESO 121-SC03}

The LMC has an unique cluster formation history
in that nearly all of its globular clusters (GCs) were formed either $\sim$ 13
Gyr ago or less than $\sim$ 3 Gyr ago (Da Costa 1991).
Recently Mackey et al. (2006) have revealed  that
the GC ESO 121-SC03 in the LMC
has ${\rm [Fe/H]}  = -0.94 \pm 0.10$ and the age range of
$8.3-9.8$ Gyr and thus confirmed
that   ESO 121-SC03 is
the only known cluster to lie squarely within the LMC age gap 
(e.g. See also Mateo et al. 1986; Geisler et al. 1997)
Previous numerical simulations showed that
the first close encounter between the LMC 
and the SMC about $3-4$ Gyr ago was
the beginning of a period of strong
tidal interaction that likely 
induced dramatic gas cloud collisions, 
leading to an enhancement of the formation of
globular clusters that has been sustained by strong 
tidal interactions to the present day (Bekki et al. 2004b).
They thus suggested that the origin of the ``age gap''
is closely associated with this dynamical coupling between
the LMC and the SMC about $3-4$ Gyr ago.
However no theoretical studies have explained
why ESO 121-SC03 is the only GC that was formed 
between $3-13$ Gyr in the history of the LMC.

The simulated small yet non-negligible amount of
the SMC's halo and old stellar components transferred into the LMC
implies that some of the SMC's old clusters, which are observed to be
significantly younger than the LMC's counterparts (e.g., Piatti et al. 2005),
can be transferred into the LMC's halo and thus identified
later as the LMC's  GC.
The pericenter distance of the LMC-SMC orbit about 0.2 Gyr 
is much smaller than that ($\sim 20$ kpc,
which is larger than the tidal radius of the SMC)  about 1.5 Gyr ago 
when the Magellanic Stream was formed during the pericenter
passage of the SMC (GN95).
It is therefore likely that GCs of the SMC
can be transferred into the LMC not at 1.5 Gyr ago but at 0.2 Gyr ago.
We accordingly suggest that the ESO 121-SC03
came from the SMC about 0.2 Gyr ago during
the last Magellanic interaction.

Previous simulations (e.g., Weinberg 2000; BC05)
suggested that the origin of the stellar halo of the LMC
is closely associated with the tidal interaction
between the LMC and the Galaxy (and the SMC).
Recent observations on kinematics of carbon stars in
the entire region of the LMC
have suggested that the LMC's stellar components
are strongly influenced by the interaction
(e.g., Olsen \& Massey 2007).
Thus the formation process of the halo GC (ESO 121-SC03)
suggested in the present paper can be  in a striking contrast
to that of the stellar halo in the LMC.

\subsection{Giant HI holes in the LMC}

Although it is well known that
gas-rich dwarf irregular galaxies (e.g., LMC, SMC, HoII, and DDO 50)
have giant HI holes in their interstellar medium,
it remains unclear what physical mechanisms are responsible
for the formation of giant HI holes in these galaxies (Rhode et al. 1999).
One of physical mechanisms for the formation of
giant HI holes (and surrounding peculiar stellar associations)
is considered to be strong impact of high-velocity clouds
onto gas disks in these galaxies (e.g., Efremov 2004 for 
detailed discussions).
Although this HI impact model can be the most natural
explanation for the LMC's HI holes with no optical counterparts
(e.g., star clusters that can form the holes through
energetic stellar winds of massive stars)(Efremov 2004),
it is unclear how such high-velocity impact of HI gas
clouds can happen in the gas disk of the LMC.

The present simulations showed that only a  small fraction
of gas (up to 8\%) initially in the SMC can be transferred
into the LMC's gas disk during the LMC-SMC interaction.
We thus propose the following scenario for the formation of 
giant HI holes in the LMC.
Owing to the large relative velocity of the Clouds ($>60$ km s$^{-1}$),
the SMC's gas clouds can collide with the LMC's gas disk
with a relative velocity higher than the sound velocities
of the gas ($\sim 4$ km s$^{-1}$).
These collisions can induce heating of gas 
in the shocked gas layers
to cause ionization of HI gas in the LMC.
The locations where
SMC's HI gas collide with the LMC
(thus where HI gas is ionized) 
can be observed as giant HI holes in the LMC.
The present simulations do not have enough resolution 
(i.e., sub-pc scale) to investigate these processes during
the LMC-SMC interaction. 
It is thus our future study to investigate
whether HI holes in the LMC can be formed from the high velocity
impact of the SMC's gas clouds 
based on more sophisticated,  high-resolution hydrodynamical
simulations of the LMC-SMC interaction.

\section{Conclusions}
We have numerically investigated stellar and gaseous distributions
and recent star formation histories of the LMC and the SMC 
interacting with each other and with the Galaxy.
We summarize our principle result as follows.

(1) The last dynamical and hydrodynamical
interaction  between the Clouds about $\sim 0.2$
Gyr ago can form the apparently
off-center bar and  peculiar  HI spirals of the LMC.
The location and the morphology of the simulated spiral
in the southern part of the LMC are similar to those
of the ``molecular ridge'' observed in the LMC
(e.g., Fukui et al. 1999; Kawamura et al. 2006).

(2) The present spatial distributions of young stars with ages
less than $\sim 20$  Myr in the LMC can be
significantly asymmetric and clumpy
owing to the interaction between the Clouds and the Galaxy.
The simulated new stars  show a strong concentration
along the LMC's stellar bar composed of old stars.
These results suggest that the observed
irregular and clumpy distributions of young stars
and star-forming regions (e.g., HII regions) in the LMC
are due to the last Magellanic interaction.

(3) A small but non-negligible fraction of stellar and
gaseous components can be transferred from the SMC into the LMC
during the interaction to form diffuse halo components around the LMC.
This result implies that some fraction of GCs in the SMC formed before the last
Magellanic interaction could have been transferred into the LMC.
We have suggested that the LMC's GC ESO 121-SC03, which
falling in the ``age-gap'' between $3-13$ Gyr, originate
from the SMC.

(4) The outer part of the SMC's gas disk can be strongly
compressed by the tidal force from the LMC and the Galaxy
and finally stretched to form tidal tails, one of which
can be observed as the MB.
During the formation of the MB,
the gas density along the forming MB exceeds the threshold gas
density of star formation so that new stars can be formed
in the MB. Star formation by the tidal compression
can explain why young stars and molecular clouds 
can be observed in the MB (e.g., Mizuno et al. 2006).

(5) The metallicity distribution function of new stars
in the MB has a peak of ${\rm [Fe/H]} \sim -0.8$, which is significantly
smaller than the central stellar metallicity of the simulated SMC 
(${\rm [Fe/H] \sim -0.6}$). 
Although the present model explains the presence of
young, metal-poor stars in the MB,
it is still unclear why some of the inter-Cloud
stars have metallicities as low as ${\rm [Fe/H] \sim -1.1}$
(e.g., Rolleston et al. 1999).

\section*{Acknowledgments} 
We are  grateful to the anonymous referee for valuable comments,
which contribute to improve the present paper. 
The numerical simulations reported here were carried out on GRAPE
systems kindly made available by the Astronomical Data Analysis
Center (ADAC) at National Astronomical Observatory of Japan (NAOJ).
K.B. acknowledges financial support from the Australian
Research Council (ARC) and ADAC throughout the course of this work.


\end{document}